\documentclass[prd,nofootinbib]{revtex4-1}
\usepackage{amsmath,amssymb}
\usepackage{slashed,verbatim,graphicx}
\usepackage{color}
\usepackage{hyperref}

\newcommand{\mt}[1]{\textrm{\tiny #1}}
\newcommand{\ri}{\mt{isco}}

\newcommand{\beq}{\begin{equation}}
\newcommand{\beqs}{\begin{equation*}}
\newcommand{\eeq}{\end{equation}}
\newcommand{\eeqs}{\end{equation*}}

\begin{document}
\setlength{\unitlength}{1mm}
\title{ISCOs in AdS/CFT}

\author{David Berenstein$^\dagger$, Ziyi Li $^\dagger$, Joan Sim\'on $^\ddagger$}
\affiliation { $^\dagger$ Department of Physics, University of California at Santa Barbara, CA 93106\\
$^\ddagger$ School of Mathematics and Maxwell Institute for Mathematical Sciences,\\
	University of Edinburgh, Edinburgh EH9 3FD, UK}

\begin{abstract} 
We study stable circular orbits in spherically symmetric AdS black holes in various dimensions and their limiting innermost stable circular orbits (ISCOs). We provide analytic expressions for their size, angular velocity and angular momentum in a large black hole mass regime. The dual interpretation is in terms of meta-stable states not thermalising in typical thermal scales and whose existence is due to non-perturbative effects on the spatial curvature. Our calculations reproduce the binding energy known in the literature, but also include a binding energy in the radial fluctuations corresponding to near circular trajectories.
We also describe how particles are placed on these orbits from integrated operators on the 
boundary: they tunnel inside in a way that can be computed from both complex geodesics in the black hole background and from the WKB approximation of the wave equation. We explain how these two 
computations are related.

\end{abstract}

\maketitle

\section{Introduction }
\label{S:Introduction}

The AdS/CFT correspondence \cite{Maldacena:1997re} is an equivalence between two seemingly distinct theories : a gauge field theory in $d$ dimensions and a gravity theory, usually realized as a string theory, on an asymptotically $AdS_{d+1}$ spacetime. The nature of the duality involves an equivalence of Hilbert spaces and a dictionary of observables from one theory to the other.

The nature of the dictionary maps any phenomenon on one side of the duality to a corresponding realization on the other side. A simple example is that states in both Hilbert spaces fit into representations of the conformal group, the latter acting as the isometry group on the gravity side. A more striking one is unitarity in the quantum field theory predicting that quantum gravity in $AdS_{d+1}$ must be unitary.

The purpose of this paper is to understand what are the consequences for the dual theory that follow from the 
existence of stable geodesic orbits in spherically symmetric $AdS_{d+1}$ black holes for $d\geq 2$. 


In the gravity side, we consider circular orbits. Classical stability requires the size of the orbit to be above of the innermost stable circular orbit (ISCO) size. We find there are no such orbits for $d=2$, recovering the result in \cite{Cruz:1994ir}, whereas for $d\geq 3$, their angular momentum $\ell$ must be above the threshold set by $\ell_\mt{isco}$, the angular momentum of the ISCO. Below this threshold, orbits plunge into the black hole.

In the field theory side, plunging orbits are associated with thermalization dynamics \cite{Sekino:2008he,AbajoArrastia:2010yt,Balasubramanian:2011ur,Nozaki:2013wia,Shenker:2013pqa}. Hence, stable circular orbits must be associated to physics that does not thermalize on a thermal time scale. However, the existence of quantum processes such as radiation or tunneling, even if subleading, can make  these states to thermalize eventually. Thus, the field theory states mapped to the gravity excitations 
must be metastable.

From the point of view of wave equations on AdS, that metastability can be 
analyzed using a WKB approximation, an analysis that was started in the work of Festuccia and Liu 
\cite{Festuccia:2008zx}. Our work is complementary to theirs, spending a lot more effort understanding the classical dynamics of these orbits and finding limits where the ISCO can be computed.
We deepen the understanding of the dual CFT physics that may be responsible 
for these phenomena.  We also provide a way to place particles in these 
trajectories from the boundary.

The kinematic conditions determining the size $(r_\mt{isco})$ and the angular momentum $(\ell_\mt{isco})$ of stable orbits are hard to solve analytically. However, working in a large mass limit, we determine their precise scaling with the temperature of the black hole in any dimension. Such scaling was also observed
in  \cite{Festuccia:2008zx}.


We argue that the origin of these meta-stable states in the dual theory are non-perturbative effects in the spatial curvature where the field theory is defined. One observation supporting this claim is the fact that at very high temperatures, the physics of the dual theory should not distinguish between flat space and a sphere, up to finite volume effects. However, the scaling with $T$ of the ISCO energy and $\ell_\mt{isco}$, which sets the window for these states to exist,  is different than the standard linear in $T$ thermal excitations. This necessarily requires the existence of a further scale, the spatial curvature. Moreover, lack of thermalization for these excitations and that this failure to thermalize stops abruptly at $\ell_\mt{isco}$ can not be explained through any perturbative argument, i.e using any polynomial in momentum. It is important to stress that finite volume is not enough. Indeed, in 2d CFTs on a finite circle, these states do not exist, matching the bulk result \cite{Cruz:1994ir} and our argument due to the absence of curvature.

A further outcome of our analysis is the calculation of the binding energy of the stable orbits in terms of conserved quantities that can be measured at the boundary. The latter decreases as we increase the angular momentum. When this binding energy is small, we can think of it as giving small contributions to the dimension of a composite operator \cite{Fitzpatrick:2014vua}. We further compute the binding energy in the radial fluctuations by including the excitations in near circular orbits. The latter is also small at large angular momentum. Both these corrections are absent in free field theories where all conformal dimensions are half-integer. Any universality that one might want to ascribe to these effects needs to be tempered with usual caveats about field theories with a gap in anomalous dimensions and where one can argue that the contributions from the gravity sector in the bulk, the stress tensor multiplet in the boundary, can dominate the result 
\cite{Heemskerk:2009pn,Hartman:2014oaa,Fitzpatrick:2014vua,Fitzpatrick:2015zha}. Results from the bootstrap at large angular momentum can  show consistency
with these ideas, at least to the extent that it shows an expansion in inverse powers of the angular momentum \cite{Fitzpatrick:2012yx,Komargodski:2012ek,Alday:2015eya}, although the analysis is usually done with four point functions of identical operators of small dimension, and the leading inverse power in the angular momentum is different: it is controlled by the twist of the stress energy tensor multiplet.

We also provide an operational way to prepare the bulk particles in these orbits by the insertion of operators in the boundary theory, extending the previous work \cite{Berenstein:2019tcs} to the black hole case. This mechanism is best understood as a tunneling process starting from the boundary. We show how the particle tunneling from the boundary on an Euclidean geodesic and the WKB calculation are related to each other using separation of variables in the Hamilton-Jacobi formulation of the geodesics. When we have angular momentum in the problem, these are complex geodesics.


The paper is organized as follows. In section \ref{sec:bulk}, we discuss gravitational physics. An exact analysis on existence, stability, angular velocity and energy of circular orbits is provided in section \ref{sec:exact}. The large mass scaling is studied in section \ref{sec:large-m} and the calculation of the radial frequency for near-circular orbits is given in section \ref{sec:near-circular}. The dual interpretation is given in section \ref{sec:cft-dual}. This contains our arguments for the relevance of spatial curvature, the range of quantum numbers for which these states exist in section \ref{sec:q-number} and also on the relation between our work and the bootstrap approach. A discussion on how to operationally define these states in the boundary theory by the insertion of boundary operators together with its bulk interpretation in terms of tunneling is addressed in section \ref{sec:operational}. We conclude with some summary of results and outlook of future directions in \ref{sec:outlook}. Appendix \ref{sec:optics} contains a more detailed discussion on some of the arguments given in section \ref{sec:operational}.

\section{Circular timelike geodesics for AdS Schwarzschild  black holes}
\label{sec:bulk}

The (d+1)-dimensional global AdS Schwarzschild black hole metric 
\begin{equation}
ds^2= -H(r)\,dt^2+ H^{-1}(r)\,dr^2 +r^2 d\Omega^2
\label{eq:AdsSch}
\end{equation}
involves a redshift factor
\begin{equation}
  H(r)= 1+\frac{r^2}{L^2}-\frac{2M}{r^{d-2}}
\label{eq:redshift}
\end{equation}
depending on the radius of AdS $L$ and the parameter $M$, related to the mass of the black hole $M_\mt{BH}$ by
\begin{equation}
  M_{\mt{BH}} = \frac{(d-1)}{16\pi G}\omega_{d-1}(2M)\,.
\label{eq:BH-mass}
\end{equation}
Here $\omega_{d-1}$ stands for the volume of the (d-1)-sphere in \eqref{eq:AdsSch}. The black hole outer horizon $r_\mt{h}$ satisfies $H(r_\mt{h})=0$ and is related to the black hole temperature $(T)$ by
\begin{equation}
    T = \frac{d}{2L^2}\,r_\mt{h}+\frac{d-2}{2r_\mt{h}}\,.
\end{equation}

The case $d=2$, corresponding to a BTZ black hole \cite{Banados:1992wn} if $M>1/2$, is special because it allows an analytic solution to $H(r_\mt{h})=0$. This can also be achieved, for any $d$, in the large mass limit, keeping $L$ fixed, when the redshift factor can be approximated by 
\begin{equation}
    H(r)\approx \frac{r^2}{L^2}-\frac{2M}{r^{d-2}}\,,
\end{equation} 
allowing to find an explicit analytic expression for $r_\mt{h}$ and its relation to the temperature \begin{equation}
  \frac{r_\mt{h}}{L}\approx \left(\frac{2M}{L^{d-2}}\right)^{1/d}\,,\quad T\,L\approx \frac{d}{2}\frac{r_\mt{h}}{L}
\label{eq:largeM}    
\end{equation}
The case $d=3$ is a black hole in $AdS_4$. The generic structure of geodesics in this spacetime has been analyzed in \cite{Cruz:2004ts}. We generalize these to higher dimensions and pay special attention to the large mass limit where additional analytic results can be found.

\subsection{Classical stability and ISCOs}
\label{sec:exact}

To study the classical stability of timelike circular orbits in these black holes, consider the geodesic quadratic action
\begin{equation}
S= \frac12\int ds\, \left(-H\,\dot t ^2+ H^{-1}\,\dot r^2 + r^2 \dot \phi ^2\right)\,,
\label{eq:geodesic-action}
\end{equation}
together with the constraint equation
\begin{equation}
-1 = -H\,\dot t ^2+ H^{-1}\,\dot r^2 + r^2 \dot \phi ^2 = g\left(\frac{d}{ds},\frac{d}{ds} \right).
\label{eq:constraint}
\end{equation}
The latter guarantees the action is proportional to the proper time $s$. Notice we already used  spherical symmetry to reduce the dynamics to an effective 3d problem \eqref{eq:geodesic-action}. Indeed, spherical symmetry guarantees any geodesic takes place in a plane of rotation where $\phi$ describes the evolution of the angular variable in such plane. 

The black hole metric \eqref{eq:AdsSch} is invariant under time translations and rotations. These give rise to two conserved quantities (on top of the constraint)
\begin{equation}
e= H(r)\,\dot t, \quad\ell = r^2 \dot \phi\,,
\label{eq:conserved}
\end{equation}
corresponding to the energy and angular momentum per unit mass, respectively. This is because our action \eqref{eq:geodesic-action} has no dependence on the mass $m$ of the particle. Plugging this back into \eqref{eq:constraint}
\begin{equation}
-1 = (-e^2+\dot r^2) H^{-1}(r) + \frac {\ell^2}{r^2}\,,
\end{equation}
the latter can be arranged into
\begin{equation}
e^2 = \dot r^2 + V(r)
\label{eq:constr}
\end{equation}
where $V(r)$ is an effective potential given by
\begin{equation}
V(r) = \left(1+\frac{\ell^2}{r^2} \right) H(r)= 1+\frac{\ell^2}{L^2}+ \frac{\ell^2}{r^2}+\frac{r^2}{L^2} - \frac{2M\ell^2}{r^{d}} -\frac{2M}{r^{d-2}}\,.
\label{eq:effpot}
\end{equation}

Circular orbits occur when $\dot r=0$. They sit at critical points $r_\mt{o}$ of $V$, $V'(r_\mt{o})=0$. Classical stability requires $V^{\prime\prime}(r_\mt{o}) > 0$ because a small increase in the energy $e$ can be compensated by a small change $\delta r$, while still finding a turning point in the constraint equation \eqref{eq:constr}. The notion of innermost stable circular orbit (ISCO) $r_\mt{isco}$ corresponds to the marginal stable orbits satisfying $V^{\prime\prime}(r_\mt{isco})=0$. Due to the change in sign $V^{\prime\prime}(r_\mt{p}) <0$ for $r_\mt{p} < r_\mt{isco}$, all such orbits would necessarily plunge into the black hole.

\paragraph{Global AdS.} It is convenient to analyze circular orbits in global AdS $(M=0)$ first. This is because, at fixed mass $M$, large circular orbits should approach the ones existing in global AdS. In this case, critical points $V^\prime(r_\mt{o})=0$ occur when
\begin{equation}
- 2 \frac{\ell^2}{r_\mt{o}^3}+2\frac{r_\mt{o}}{L^2}=0 \quad \Rightarrow \quad r_\mt{o}= \sqrt{\ell\,L}\,, \,\, e= \frac{\ell}{L} +1
\label{eq:vac-scaling}
\end{equation}
where we used \eqref{eq:constr} and \eqref{eq:effpot} for $M=0$. Since $V^{\prime\prime}(r_\mt{o}) > 0$ for all values of the angular momentum $\ell$, all circular orbits are classically stable. Hence, there is no ISCO in global AdS. The particle angular velocity in global time equals
\begin{equation}
  \Omega = \frac{d\phi}{dt} = \frac{\ell}{r_\mt{o}^2}\frac{r_\mt{o}^2/L^2+1}{e} =\frac{1}{L}\,.
\label{eq:ang-vel-global}
\end{equation}
This equals $\omega= \frac{d e}{d\ell}$ and corresponds to propagation at the speed of light in the boundary of AdS. To sum up, there exists a single circular orbit for \emph{any} value of the angular momentum $\ell$ related to the size of the orbit $r_\mt{o}$ by the scaling relation \eqref{eq:vac-scaling}.

\paragraph{Schwarzschild AdS black holes.} Let us consider stable circular orbits in the black holes \eqref{eq:AdsSch}. Criticality imposes the condition
\begin{equation}
 V^\prime(r_\mt{o}) = -2 \frac{\ell^2}{r_\mt{o}^3}+2\frac{r_\mt{o}}{L^2} + d \frac{2M\ell^2}{r\mt{o}^{d+1}} +(d-2)\frac{2M}{r_\mt{o}^{d-1}}=0\,. 
\label{eq:dv}
\end{equation}
Solving for $\ell(r_\mt{o})$, we get
\begin{equation}
  \ell^2= \frac{r_\mt{o}^4}{L^2}\frac{(r_\mt{o}^d+(d-2)M\,L^2)}{r_\mt{o}^d- d M r_\mt{o}^2}.
\label{eq:l-rad}
\end{equation}
As expected, when $r_\mt{o}\to \infty$ and $d>2$, at fixed $M$, we recover the circular orbits of global AdS \eqref{eq:vac-scaling}, where $r_\mt{o}\simeq \sqrt{\ell\,L}$. In fact, for $d>2$, $\ell^2\,L^2 > r_\mt{o}^4$. This means circular orbits get pushed in relative to global AdS. Equivalently, the radius of the orbit at fixed angular momentum shrinks. Even though the current discussion is coordinate dependent, this conclusion reflects the attractive nature of gravity. The $d=2$ BTZ black hole behaves differently. Notice $\ell^2$ is always negative whenever $M>1/2$. Thus, there are {\em no circular geodesic orbits} in BTZ, reproducing a well known statement \cite{Cruz:1994ir}.  

Circular orbits stop existing at the radius $r_\mt{min}$, defined by 
\begin{equation}
  r_\mt{min}^d-d Mr_\mt{min}^2=0 \quad \Rightarrow \quad r^{d-2}_\mt{min} = dM\,.
\label{eq:rmin}
\end{equation} 
This is because $\ell^2$ is negative for $r_\mt{o} < r_\mt{min}$, whereas $\ell^2$ is large for $r_\mt{o} \gtrsim r_\mt{min}$.  A better understanding of this geometric locus can be gained in terms of a limiting set of light-like geodesics. Indeed, the constraint equation for light-like geodesics
\begin{equation}
(-e^2+\dot r^2) H^{-1}(r) +\frac{\ell^2}{r^2} =0
\end{equation}
gives rise to the effective potential
\begin{equation}
V_\mt{l}(r) =\ell^2 \frac{H(r)}{r^2} = \ell^2\left(\frac{1}{L^2}+\frac 1{r^2} -\frac{2M}{r^d}\right)\,.
\label{eq:vlight}
\end{equation}
Circular light-like geodesics correspond to $V_\mt{l}^\prime=0$. These are independent of $\ell$ and
occur when 
\begin{equation}
  V^\prime_\mt{l}(r)= -\frac{2}{r^3}+d \frac{2M}{r^{d+1}} =0\,.
\end{equation}
or equivalently, at $r^d -d Mr^2=0$. Thus, the locus $r_\mt{min}$ corresponds to the light-ring of the black hole. Notice there is no light-ring for BTZ black holes since $V^\prime_\mt{l}(r) > 0$ for $d=2$ and $2M>1$, in agreement with our previous observation. To sum up, all circular timelike orbits, stable and unstable, must occur beyond the light-ring radius, i.e. $r_\mt{o} > r_\mt{min}$.

\paragraph{Stability.} Stability requires the second derivative of the effective potential \eqref{eq:effpot} to be positive
\begin{equation}
 r_\mt{o}\,V^{\prime\prime}(r_\mt{o}) = \frac{6\ell^2}{r_{\mt{o}}^3} + \frac{2r_\mt{o}}{L^2}-d(d+1)\frac{2M\,\ell^2}{r_\mt{o}^{d+1}}-(d-2)(d-1)\frac{2M}{r_\mt{o}^{d-1}}> 0\,,
\label{eq:ddV} 
\end{equation}
together with the critical condition \eqref{eq:dv}. We discuss the notion of marginal stability, i.e. the one algebraically characterised by $V^{\prime\prime}(r_\mt{isco})=0$, first.

Solving \eqref{eq:dv} and $r_\mt{isco}\,V^{\prime\prime}(r_\mt{isco})=0$ for $r_\mt{isco}/L^2$ and equating the corresponding expressions, we derive the constraint
\begin{equation}
    8\ell_\mt{isco}^2\,r_\mt{isco}^{d-2} = d(d+2)\,2M\,\ell_\mt{isco}^2 + d(d-2)\,2M\,r_\mt{isco}^2
\label{eq:isco-cons}
\end{equation}
This relates the locus of the orbit $r_\mt{isco}$ with its angular momentum $\ell_\mt{isco}$. When the latter holds, one must still solve a second constraint, say $V^\prime(r_\mt{isco})=0$. 

When plotting or numerically studying these two conditions, one finds that classical stability requires $\ell > \ell_\mt{isco}$, with $\ell_\mt{isco}$ satisfying a non-trivial algebraic condition. This statement can be made explicit for $d=4$. In this particular case, \eqref{eq:isco-cons} becomes a quadratic equation that is solved by
\begin{equation}
  r_\mt{isco}^2 = 3\frac{2M\,\ell_\mt{isco}^2}{\ell_\mt{isco}^2-2M}\,.
\end{equation}
Obviously, this requires $\ell_\mt{isco}^2 > 2M$. Furthermore, plugging this back into \eqref{eq:dv}, we obtain a second condition for $\ell_\mt{isco}$
\begin{equation}
  L^2(\ell_\mt{isco}^2-2M)^3 = 27\,(2M\ell_\mt{isco}^2)^2\,.
\end{equation}
This condition was already derived in \cite{Festuccia:2008zx}. The lesson of this discussion is that for $d>2$ there exists a critical angular momentum $\ell_\mt{isco}$, the one corresponding to the marginal orbit $r_\mt{isco}$, such that for $\ell > \ell_\mt{isco}$, there exist classical stable circular orbits in the AdS black holes \eqref{eq:AdsSch2}.

\paragraph{Angular velocity and energy.}  The angular velocity $\Omega$ of the stable orbits equals
\begin{equation}
  \Omega= \frac{\dot \phi}{\dot t} = \frac{\ell}{r_\mt{o}^2}\frac{H(r_\mt{o})}{e}\,.
\end{equation} 
This is equivalent to the definition
\begin{equation}
  \omega= \frac{d e}{d \ell}\,,
\end{equation}
because for circular orbits the angular momentum depends on the size of the orbit through \eqref{eq:l-rad}. Hence, the constraint equation \eqref{eq:constr} becomes
\begin{equation}
  e^2= V(\ell,r_\mt{o})=\left(\frac{\ell^2}{r_\mt{o}^2}+1\right)H(r_\mt{o})\,.
\label{eq:concirc}
\end{equation}
Taking the total derivative with respect to $r$, we find that
\begin{equation}
2e\frac{de }{dr} = \partial_\ell V \frac{d\ell}{dr} +\partial_r V= \partial_\ell V \frac{d\ell}{dr}
\end{equation} 
From this expression, we derive the equivalence 
\begin{equation}
  \omega=\left(\frac{de }{dr}\right)\left(\frac{d\ell}{dr}\right)^{-1}= \frac{de}{d\ell}=\frac{\partial_{\ell}V}{2e} = \frac{\ell}{r_\mt{o}^2} \frac{H(r_\mt{o})}{e}=\Omega
\label{eq:ang-vel}
\end{equation}
Using the identity
\begin{equation}
   1+\frac{\ell}{r_\mt{o}^2} = \frac{H}{1-dM/r_\mt{o}^{d-2}}
\end{equation}
in \eqref{eq:concirc} and plugging the corresponding energy in \eqref{eq:ang-vel}, we obtain an exact expression for the angular velocity of circular stable orbits
\begin{equation}
  \omega\,L= \sqrt{1+(d-2)\frac{ML^2}{r_\mt{o}^{d}}}\,.
\label{eq:ome}
\end{equation}
At fixed $M$ and large orbit size, $\omega\,L \to 1$ reproducing the expected global AdS result \eqref{eq:ang-vel-global}. Corrections to this result
\begin{equation}
\frac{de}{d\ell} \simeq \frac{1}{L}\sqrt{1+\frac{(d-2)ML^2}{(L\ell)^{d/2}}}\simeq \frac{1}{L}\left(1 +\frac{d-2}{2} \frac{ML^2}{(L\ell)^{d/2}}\right),
\end{equation}
where we used $r_\mt{o}\approx (\ell\,L)^{1/2}$, allow us to determine the binding energy of this large size orbits by integrating with respect to $\ell$
\begin{equation}
e(\ell)= \int^\ell {\frac{de}{d\ell}} \simeq \frac{\ell}{L} +1 -\frac{M}{(L\ell)^{d/2-1}}\,. 
\label{eq:envsj}
\end{equation}
Notice the arbitrary constant was fixed by matching the global AdS result \eqref{eq:vac-scaling}. The negative term provides a gauge invariant characterisation of the binding energy of the circular orbit, i.e. independent of the choice of radial coordinate $r$. That the binding energy is negative indicates the attractive nature of gravity\footnote{The correct binding energy requires to multiply our correction by the particle mass $m$ due to our choice of classical action \eqref{eq:geodesic-action}.}. Our result is consistent with the binding energy found in \cite{Fitzpatrick:2014vua} (see also \cite{Kusuki:2018wpa}). 
Additional corrections will arise from both expanding the square root and remembering that $r(\ell) <\sqrt{\ell\,L}$ receives corrections from large $\ell$.

Once more, the $d=2$ case is special. The angular velocity $\omega\,L=1$ everywhere, and not just close to the boundary. Hence, the binding energy must be independent of the orbit size. Indeed, solving \eqref{eq:l-rad} for $r_\mt{o}/L$ and plugging this into the energy \eqref{eq:concirc}, we find
\begin{equation}
e(\ell)= \sqrt{1-2M} +\frac{\ell}{L}\,, \quad \quad M < \frac{1}{2}\,.
\end{equation}
This is the energy of the orbit in an AdS$_3$ conical defect. As soon as $M>1/2$, the energy becomes imaginary, indicating the corresponding circular orbits become unstable, as we already established.

Observe that some analogue of Kepler's third law, i.e. a simple scaling relation between the size of the orbit and the period, would exist if the second term in \eqref{eq:ome} would be dominant. Since stable orbits have size $r_\mt{o} > r_\mt{min}$, this requires a small $M$ and $r_\mt{o}$ limit, i.e. $L\gg r_\mt{o}$, while satisfying $r_\mt{o}^{d-2} > dM$. In this limit, we can approximate \eqref{eq:l-rad} by
\begin{equation}
  \frac{\ell^2}{r^3_\mt{o}} \approx (d-2)\frac{M}{r^{d-1}_\mt{o}}\,\frac{1}{1-dM/r^{d-2}_\mt{o}}\,.
\end{equation}
Plugging this back into \eqref{eq:ddV} and after some algebra, we find
\begin{equation}
  r_\mt{o}\,V^{\prime\prime}(r_\mt{o}) \approx 2(d-2)\frac{M/r^{d-1}_\mt{o}}{1-dM/r^{d-2}_\mt{o}}\left(2-d + 2\left(1-d\frac{M}{r^{d-2}_\mt{o}}\right)\right) + 2\frac{r_\mt{o}}{L^2}
\end{equation}
We learn these orbits are not stable for $d\geq 4$, but they are stable for $d=3$ if $3M/r_\mt{o} < 1/2$, a situation where the frequency is approximately
\begin{equation}
  \omega \approx \sqrt{3M}\,r^{-3/2}_\mt{o}\,,
\end{equation}
the same scaling in Kepler's law. 
In this case $r_\mt{o}\ll L$, so these orbits do not really feel the curvature of $AdS_4$ and can be considered as circular orbits in flat space Schwarzschild. These are known to satisfy Kepler's third law in the standard Schwarzschild coordinate system where the radius of the sphere $d\Omega^2$ determines the radial coordinate. This is exactly our choice in equation \eqref{eq:AdsSch}.

\subsection{Large mass limit} 
\label{sec:large-m}

We established all circular orbits satisfy $r_\mt{o} > r_\mt{min}$, whereas the classically stable ones occur for $r_\mt{o} > r_\mt{isco}$, with the marginal orbit $r_\mt{isco}$ determined by solving the kinematic constraints \eqref{eq:isco-cons} and \eqref{eq:dv}. 

These constraints can not be solved analytically, unless we consider a large black hole mass limit. Since $r_\mt{isco} > r_\mt{min}$ and $r_\mt{min} \sim M^{1/(d-2)}$ for any mass, it is natural to consider a large $M$ limit keeping the locus of the black hole light-ring $M/r^{d-2}$ fixed. Notice this scaling is different from the one determining the horizon in \eqref{eq:largeM}.

To further appreciate the relevance of this physical scaling, introduce the dimensionless coordinate $\hat{r}=r/L$ and parameter $\hat{M}=M/L^{d-2}$ and further plug \eqref{eq:l-rad} into \eqref{eq:ddV}. After some algebra, this can be written as
\begin{equation}
    V^{\prime\prime} = \frac{1}{L^2}\frac{1}{1-d\frac{\hat{M}}{\hat{r}^{d-2}}}\left[8-\frac{2\hat{M}}{\hat{r}^{d-2}}\left(d(d+2) + \frac{d-2}{\hat{r}^2}\left(d-4 + d\,\frac{2\hat{M}}{\hat{r}^{d-2}}\right)\right)\right]\,.
\label{eq:interm}
\end{equation}
When we consider the large $M$ limit keeping $M/r^{d-2}$ fixed, the value of $\hat r$ is large and we can approximate the answer by
\begin{equation}
    V^{\prime\prime} \approx \frac{2}{L^2}\frac{1}{1-d\frac{\hat{M}}{\hat{r}^{d-2}}}\left[4-\frac{\hat{M}}{\hat{r}^{d-2}}\left(d(d+2) \right)\right]\,.
\label{eq:interm2}
\end{equation}
The marginal orbits $(V^{\prime\prime}(r_\ri)=0)$ are located at
\begin{equation}
  r_{\ri} \approx \left(\frac{d(d+2) M}{4}\right)^{1/(d-2)}
\label{eq:isco}
\end{equation}
and have angular momentum
\begin{equation}
  \frac{\ell_{\ri}}{L} \approx \sqrt{\frac{d+2}{d-2}} \hat{r}^2_\mt{isco}\,.
\end{equation}
This large mass limit also allows us to compare the ISCO size with the horizon size \eqref{eq:largeM}. Indeed,
\begin{equation}
  \frac{r_\mt{isco}}{L} \approx \left(\frac{d(d+2)}{8}\right)^{1/(d-2)}\,\left(\frac{r_\mt{h}}{L}\right)^{1+2/(d-2)}\,.
\end{equation}

Using \eqref{eq:interm}, classical stability reduces to $r>r_{\ri}$. This is because whenever the numerator in \eqref{eq:interm} is positive, i.e. $r>r_{\ri}$, the denominator factor is also positive for $d>2$. 
Furthermore, as stressed earlier, the angular momentum of stable orbits must indeed be above the marginal $\ell^2_{\ri}$. Indeed, the difference
\begin{equation}
  \frac{\ell^2}{L^2} - \frac{\ell^2_{\ri}}{L^2} = \frac{\hat{r}^4}{1-\frac{4}{d+2}a}\left[1-\frac{d+2}{d-2}\,a^4 + \frac{4}{d-2}\,a^{d+2}\right] \quad \text{with} \quad a\equiv \frac{r_{\ri}}{r}
\end{equation}
is a monotonically decreasing function $\forall\,a\in (0,1)$.

In general, it is not analytically possible to obtain $e(\ell)$ since this requires to invert \eqref{eq:l-rad} in order to write $r_\mt{o}(\ell)$. This is because the powers of $M/r_\mt{o}^{d-2}$ appearing in \eqref{eq:l-rad} are kept fixed in the large mass limit, making the analytic inversion not possible. 


What we can determine is the scaling with the black hole temperature of both the energy and angular momentum of these orbits. According to \eqref{eq:largeM}, the temperature of the black hole in the large mass limit is
\begin{equation}
  T\,L \approx \frac{d}{2}\,(2\hat{M})^{1/d}\,.
\end{equation}
Hence, the energy and angular momentum of the classically stable circular orbits scale with the temperature as
\begin{equation}
  e\,,\frac{\ell}{L} \simeq (T\,L)^{2d/(d-2)}\,.
\label{eq:e-ell-largeM}
\end{equation}
Notice that since $2d/(d-2)>1$, the power of the temperature in these expressions indicates a scaling that increases faster than $T$ for the stable orbits.

\subsection{Near-circular orbits}
\label{sec:near-circular}

We can extend our analysis to near-circular orbits by including radial excitations at fixed $\ell$. The dynamics of these orbits can be understood by expanding the constraint equation \eqref{eq:constr} around a circular orbit using $r= r_{\circ}+\delta r$ and $e=e_{\circ}+\delta e$. The constraint becomes
\begin{equation}
2e_\circ\,\delta e= \left(\frac{d\delta r}{dt}\right)^2\,\dot{t}^2+ \frac{1}{2} V^{\prime\prime}(r_{\circ}) \delta r^2
\label{eq:q-harmonic}
\end{equation}
This is identical to the energy of a harmonic oscillator with mass $m=\dot{t}^2/e_\circ$ and spring constant $k=V^{\prime\prime}(r_{\circ})/(2e_\circ)$. It follows the angular frequency $\omega_\mt{r}$ of these radial oscillations is given by
\begin{equation}
\omega_r^2=\frac{k}{m}= \frac{1}{2}V^{\prime\prime}(r_{\circ})\,\left(1-d\frac{M}{r_\circ^{d-2}}\right) = \frac{1}{L^2}\left[4-\frac{\hat{M}}{\hat{r}_\circ^{d-2}}\left(d(d+2) + \frac{d-2}{\hat{r}_\circ^2}\left(d-4 + d\,\frac{2\hat{M}}{\hat{r}_\circ^{d-2}}\right)\right)\right]\,,
\label{eq:omega-rad}
\end{equation}
where in the second equality, we used
\begin{equation}
    \dot{t} = \frac{e_\circ}{H} = \frac{1}{\sqrt{1-d\frac{M}{r_\circ^{d-2}}}}\,,
\end{equation}
and in the third we simplified using \eqref{eq:interm}. The frequency \eqref{eq:omega-rad} is exact and valid for any stable orbit $(r_\circ > r_\ri)$. 

For both large orbits at \emph{fixed} mass $M$ and the large mass limit discussed in section \ref{sec:large-m}, the frequency of the radial oscillations reduces to
\begin{equation}
    (\omega_\mt{r}\,L)^2 \approx 4 - d(d+2)\frac{M}{r_\circ^{d-2}} 
\end{equation}
For large orbits at \emph{fixed} mass $M$, specifically, $r_\circ\simeq \sqrt{\ell\,L}$, we can rewrite this frequency as
\begin{equation}
  \omega_\mt{r}\,L \simeq 2- \frac{d(d+2)}{4}\frac{M}{(\ell\,L)^{d/2-1}}\quad \quad (d\geq 3)\,.
\label{eq:omegar-cor}
\end{equation}
The case d=2 is, once more, special. From the exact expression \eqref{eq:omega-rad}, we learn
\begin{equation}
  \omega_\mt{r}\,L = 2\,\sqrt{1-2M}\,.
\end{equation}
This is $\ell$ independent and it vanishes at the threshold of black hole formation, i.e. $M=1/2$. The frequency becomes imaginary for $M>1/2$, in agreement with these orbits being unstable, and it decreases from $\omega_\mt{r}\,L=2$, for global AdS$_3$, to zero, as the conical defect mass $M$ increases.

\section{The CFT dual interpretation} 
\label{sec:cft-dual}

To discuss the dual field theory interpretation of the bulk results presented in the previous section, we first use the basic holographic dictionary mapping a bulk particle of mass $m$ to a single trace operator of dimension $\Delta \simeq m\,L+d/2$ \cite{Gubser:1998bc, Witten:1998qj}. The additional $d/2$ contribution can be thought of as a zero point energy fluctuation resulting from confinement at the origin from a gravitational potential (see for example \cite{Berenstein:2002ke}). The energy $e$ and angular momentum $\ell$ per unit mass of the bulk particle excitation translate into an energy $E$ and angular momentum $J$ in the CFT given by
\begin{equation}
  E= \Delta e\,, \quad J= \Delta \frac{\ell}{L}\,.
\label{eq:charge-dic}
\end{equation}

What is the possible meaning of the ISCO in the dual field theory? In the bulk spacetime, the ISCO defines an orbit size separating plunging orbits, i.e. those falling directly into the black hole, from classically stable circular orbits. Once we include semiclassical corrections, these orbits become meta-stable and bulk particles can decay semiclassically by tunneling past the potential barrier and plunging into the black hole geometry. These effects are suppressed by a tunneling amplitude of the form $A_{\mt{tun}} \simeq \exp( - \Delta S)$, with $S$ the tunneling action for a point particle of unit mass to overcome the potential barrier\footnote{Gravitational radiation can also cause the orbit to lose energy, but this effect is expected to be small \cite{Fitzpatrick:2014vua}.}. The semiclassical tunneling of wave solutions (which can be computed using a WKB approximation) has also been studied to discuss black hole stability \cite{Dias:2012tq} and is an important ingredient in the analysis.

There is plenty of evidence in the AdS/CFT literature indicating that the classical plunging should be interpreted as the dynamics of thermalization for typical excitations of the field theory. This is supported by the scrambling properties of black holes \cite{Sekino:2008he}, the time evolution of holographic entanglement entropy \cite{AbajoArrastia:2010yt,Balasubramanian:2011ur,Nozaki:2013wia}
and two-point functions \cite{Balasubramanian:2011ur} after CFT quench perturbations, and the butterfly velocity spread of quantum information under small perturbations near the horizon of a black hole \cite{Shenker:2013pqa}. Applying the same logic, meta-stable orbits should describe field theory long-lived excitations that do not thermalise like typical excitations\footnote{The existence of such long-lived meta-stable states was already stressed in \cite{Festuccia:2008zx} when analytically estimating the frequency of the AdS black hole quasi-normal modes in a WKB approximation.}. We are interested in understanding when these excitations appear and how their parameters scale with the temperature of the black hole. Using the AdS/CFT correspondence, the bulk kinematics presented in the previous section will provide answers to these questions. 

However, before computing the dual CFT charges $E$ and $J$ for these excitations, we would like to 
interpret the existence of these orbits as a non-perturbative curvature effect on the boundary field theory. In our case, the curvature of the (d-1)-sphere. 

To gain some intuition, let us compare our AdS black holes \eqref{eq:AdsSch} with planar AdS black holes having a flat cross section. These have infinite mass, but finite mass density. Their dual involves a CFT on \emph{flat space} at finite temperature. Their dual infinite energy is due to the infinite volume of flat space. Since this field theory has no scale on its own, the only energy scale in the thermal CFT is the temperature itself.

One way to derive these black holes from their global AdS versions in \eqref{eq:AdsSch} (see \cite{deBoer:2011zt} for example), is to combine a mass rescaling $M\to \lambda^d\,M_0$ with a conformal transformation $(M, L)\to (\lambda^{-1}\,M,\,\lambda\,L)$, in the limit $\lambda\to \infty$ so that
\begin{equation}
  (M,\,L) \to (\lambda^{d-1}\,M_0,\,\lambda\,L) \quad \quad \lambda\to \infty
\end{equation}
Combining this transformation with the coordinate rescaling $r=\lambda^{(d+1)/d}\,\rho$, the metric \eqref{eq:AdsSch} becomes
\begin{equation}
ds^2= -dt^2\lambda^{2/d} (\rho^2/L^2-2M_0\rho^{2-d}) + d\rho^2 (\rho^2/L^2 -2M_0\rho^{2-d})^{-1} +\lambda^{2(d+1)/d} \rho^2 d\Omega^2 .
\end{equation}
To take the limit $\lambda\to\infty$ requires the time coordinate rescaling $\tilde t = \lambda^{1/d} t$ and approximating the sphere metric by its tangent plane at a point. To see the latter, expand the sphere metric around the north pole in a flat coordinate system, $d\Omega^2= d\vec \theta^2 + a\, \vec \theta^2 d\vec \theta^2+\dots $ and define a new rescaled angular variable $\vec  x_\perp= \lambda^{(d+1)/d}\,\vec \theta$. The sphere metric then becomes $d\Omega^2= \lambda^{-2(d+1)/d} ( d\vec {x_\perp}^2 +\mathcal{O}(a \lambda^{-2(d+1)/d}))$. Altogether, the dominant limiting $\lambda\to\infty$ metric becomes
\begin{equation}
ds^2= -d\tilde t^2 (\rho^2/L^2-2M_0\,\rho^{2-d}) + d\rho^2 (\rho^2 -2M_0\,\rho^{2-d})^{-1} + \rho^2 d\vec{x}_{\perp}^2\,.
\label{eq:AdsSch2}
\end{equation}
Besides time translation symmetry in $\tilde t$, the resulting metric is invariant under the full euclidean group acting on $\vec{x}_\perp$, in agreement with the flat cross-section of the AdS black hole that resulted from zooming into a small angular region around the north pole of the coordinate system. The location of the horizon $\rho_{\mt{h}}^d=2M_0\,L^2\equiv \gamma$, determines the scale of the energy $(\tilde{e})$ and linear momentum $(p_\perp)$ for particles in these planar AdS black holes to be
$\tilde e \sim \gamma^{-1/d} $ and $p_{\perp}\sim \gamma^{-1/d}$. Both scale with the temperature since $T\sim \gamma^{-1/d}$. Thus, the double scaling limit preserves CFT excitations scaling like
$E\simeq \Delta T$ and $J \simeq \Delta T$, the latter being converted to linear momentum in the limit. 

From the CFT point of view, black holes are dual to a finite temperature configuration.
On the sphere, there are now two scales : the radius of the sphere and the thermal length scale. However, at very high temperature, the thermal scale is very short compared to the radius of the sphere, and the main difference between the sphere and flat space is the finite volume of the sphere.
Since correlation functions decay exponentially in distance with the thermal scale, the effects of the finite volume naively seem to wash out. Equivalently, the physics of flat space and the sphere at high temperature should be almost the same, except for finite volume effects. When we zoom into a region which is of the order of the thermal scale, the natural energies and excitations that are seen depend only on the thermal scale, so that $E,p\simeq T$, exactly like the double scaling limit of the dual black hole tells us. 

By contrast, the energies and angular momentum of the stable particles in ISCO trajectories around global AdS black holes \eqref{eq:AdsSch} scale like \eqref{eq:e-ell-largeM}
\begin{equation}
e\,, \frac{\ell}{L} \simeq \hat{M}^{2/({d-2})} \simeq  (T\,L)^{\frac{2d}{d-2}}\,. 
\label{eq:scaling}
\end{equation}
This is a different scaling from the flat $\lambda\to \infty$ double scaling limit, i.e. the global ISCO energies and angular momenta grow faster than linearly in the temperature. The same scaling has been observed in \cite{Festuccia:2008zx}. This means the global ISCO trajectories get pushed out of the flat rescaled coordinate system, i.e. they are too far in the $\rho$ direction to be captured by the flat 
black hole cross section. Hence, the existence of global ISCO trajectories must be a property of the global AdS geometry \eqref{eq:AdsSch}. They disappear in the planar limit. The one distinction between these two in the boundary field theory is that the boundary geometry $S^{d-1}$ has an additional scale: the radius of curvature, which is absent in ${\mathbb R}^{d-1}$. This is precisely the scale $L$ appearing on the right hand side of \eqref{eq:scaling} to restore the units. 

The natural interpretation of this observation is that the metastability of the states associated to the stable orbits is due to a curvature effect in the field theory that competes with the temperature. The effect is absent in $3d$ black holes (BTZ black holes), even if there is still a finite volume on the boundary. As we have seen, all geodesics plunge in this case. Finite volume alone is not enough. We need the change in the curvature of the boundary. The effect does not exist either if we take a flat space black hole with periodic spatial identifications: all geodesics plunge in that case as well.

Moreover, the width of the wave functions for these states becomes exponentially small, rather than power law, in the temperature. This exponentially small behaviour is the semiclassical result of the gravity computation, which we assume gives the correct CFT description of the metastability of the perturbations. Since perturbative Feynman diagrams can only give power laws in E and J,  
we must conclude that the dynamics responsible for this effect is \emph{non-perturbative} in the curvature of the $(d-1)$-sphere, since this is the only way to obtain such an exponential suppression in E and J.

\subsection{Quantum numbers}
\label{sec:q-number}

Having clarified the origin for the metastability of these states in the dual CFT, we return to the discussion of their quantum numbers $E$ and $J$. For fixed mass and large orbits, using the dictionary \eqref{eq:charge-dic} in \eqref{eq:envsj}, stable circular orbits around the black hole should correspond to CFT excitations with energy
\begin{equation}
  E= \Delta+ J -\Delta \frac{M}{L^{d-2}} \left(\frac{\Delta}{J}\right)^{d/2-1}\,.
\label{eq:bootstrap}
\end{equation}

Before including the radial fluctuations due to near circular orbits, we discuss the relation between our result \eqref{eq:bootstrap} and the existing literature. First, notice that if we use $M = \frac{8\pi G_\mt{N}}{(d-1)\,\omega_{d-1}}\,M_\mt{BH}$, which follows from \eqref{eq:BH-mass}, we reproduce the dominant contribution to equation (2.18) in \cite{Fitzpatrick:2014vua} up to a missing $1/2$ in the last term. We believe this is due to using different conventions for $M$. This observable in \cite{Fitzpatrick:2014vua} was computed assuming that one particle (the black hole in our case) is fixed, while the other orbits with large $\ell$, precisely the bulk situation we discussed in the previous section.

Next, we compare with the Bootstrap programme computing the anomalous dimensions of composite operators $[\mathcal{O}_1\mathcal{O}_2]_\ell$ in the CFT. Bootstrap calculations require one to have operators with dimension $\Delta_{BH}+\Delta+J+\mathcal{O}(1/J^\tau) $ where $\tau$ is the twist of the stress energy tensor, $\tau=d-2$  \cite{Fitzpatrick:2012yx,Komargodski:2012ek,Alday:2015eya}. This result is valid only at asymptotically large values of the spin.
Our corrections in \eqref{eq:bootstrap} are of order $\mathcal{O}(1/J^{\tau/2})$ \footnote{In the special case $d=2$ we get the correct answer with twist zero.}. These are not in contradiction to each other. The black hole is also expected to have internal excitations with energies and spin very close to the semiclassical calculation above. These would correspond to black holes with angular momentum $J$ and energy $\Delta_{BH}+\Delta+J+\mathcal{O}(1/J^\tau) $.
They are not these states. The states we are discussing are only expected to be metastable and do not represent exact dimensions of conformal primaries. The metastable states should eventually decay to these black hole excitations, either by tunneling or some other decay process (e.g., emissions of gravitational waves).
Coming back to the analysis in \cite{Fitzpatrick:2014vua}, the validity of their equation 2.10 relating the asymptotic angular momentum $\ell_1$ of $\mathcal{O}_1$ with the total angular momentum $\ell$, requires the latter to be much larger than the sum of the dimension of the two primaries, i.e. $\ell \gg \Delta_1 + \Delta_2$. In such situation, one would indeed expect the black hole to back react and move. Notice this is \emph{not} the regime where our approximations hold, where the black hole is expected to be heavy and to have $\Delta_{BH}\gg J$. 

Finally, let us compare with the bootstrap analysis of 4-pt functions involving two heavy and two light operators, as studied in \cite{Kulaxizi:2018dxo, Karlsson:2019qfi,Karlsson:2019dbd,Li:2019zba,Li:2020dqm}, and references therein. This approach is based on an analytic continuation to the Regge limit, so that in the bulk one focuses on geodesic motion along null geodesics in the presence of a black hole. This provides a computation of the phase shift. The latter can be dominated by an exchange in the stress tensor multiplet in the $t$-channel (in the particle physics language). To understand the dimensions of composite operators in this situation, one needs to transform the $t$-channel expansion to the $s$-channel. To do so, the literature typically assumes that one can treat the system as a ``double trace" operator near a generalized free field limit. Under that assumption, one can recover the same correction $\mathcal{O}(1/J^{\tau/2})$ in the binding energy as in \eqref{eq:bootstrap}. However, there is no decay observed in these discussions since, ab initio, due to the generalised free field approximation, such possibility does not exist. A complete calculation would deal with operators whose conformal dimensions are near the bound state energy. To observe the decay would require to work with a very fine-grained density of states that captures the asymptotics of OPE coefficients. Our understanding is the approximations used in this body of work are able to estimate an average of these OPE coefficients convoluted with the density of states with the right intermediate quantum numbers. A full computation would then be able to give a prediction of the `lifetime' of the orbits. We would like to stress that our results seem to get the exact correct answers for the case of conical singularities in $AdS_3$. In these cases, since there are no black holes in the bulk, the density of states is reduced (there is very little entropy) and the double trace approximation is probably better justified, especially at large values of the angular momentum. 

We can now add the radial fluctuations associated to nearly circular orbits by treating these excitations as a harmonic oscillator with frequency $\omega_r$, derived in equation \eqref{eq:omegar-cor}. These provide an additional contribution to the energy $\delta E=k\, \omega_r L$, with $k$ the occupation number\footnote{To be more precise, the bulk analysis for near circular orbits \eqref{eq:q-harmonic} gives rise to a quantum harmonic oscillator with $\hbar = 1/m$. Hence, $\delta e = k\,\hbar \omega_\mt{r} = (k/\Delta)\,\omega_{\mt{r}}\,L$, with $k$ the occupation number. Using our AdS/CFT dictionary \eqref{eq:charge-dic}, the CFT energy $\delta E = \Delta \delta e = k\,\omega_{\mt{r}}\,L$.}, so that the total CFT energy equals
\begin{equation}
E= \Delta+ J +2k - \Delta\,\left(1 + \frac{k}{\Delta} \frac{d(d+2)}{4}\right)\,\frac{M}{L^{d-2}} \left(\frac{\Delta}{J}\right)^{d/2-1} \quad \quad (d\geq 3)
\label{eq:desc}
\end{equation}
Again, the $\delta\omega_r<0$ indicates that there is also a binding energy cost for the radial oscillations.  These energy costs for the radial direction are comparable to the total binding energy when $k\simeq \Delta$, that is,  when the radial excitation energy is comparable to the mass of the particle.  

The analogous energy for 2d CFTs equals
\begin{equation}
  E = \Delta\,\sqrt{1-2M}\,\left(1 + 2\frac{k}{\Delta}\right) + J \quad \quad (d=2)
\end{equation}

For $M=0$, we recover the usual descent relation for energies (dimensions) in the representations of the conformal group, where $E= \Delta + J +2k$ counts descendants of the operator ${\cal O}_\Delta$ of the type
$\partial_ \mu^J \square^k {\cal O}_{\Delta}$ \footnote{In \cite{Fitzpatrick:2014vua} a different argument is used to suggest 
a similar result.}. 
When we deal with a black hole where $M\neq 0$, these can not be thought of as descendants. Descendants have a difference of energy with a primary that is exactly an integer.
The fact that we see small corrections in the energy means that we need to look for the descendants of these states elsewhere. 
The descendants will be generated by the center of mass
motion of the black hole instead. If we think of the black hole as an object that leads to a spontaneous breaking of the conformal symmetry, the associated ``goldstone bosons" would be associated to collective motion of the black hole itself. The other Virasoro descendants will still be generated by boundary 
gravitons.

In a sense, the fact that in \eqref{eq:desc} we get corrections in both $J$ and $k$ from the conformal group representation theory values suggest that we should interpret these 
states as different (approximate  \footnote{These states are only metastable, so they do not describe exact dimensions of operators.})   primaries with dimension $\Delta_{tot}\simeq E_{BH}+E$, 
 The fact that the corrections in the energy are small but not zero at large $J$ means that these results do not apply to 
free field theories, where all dimensions of operators are integers or half integers. 
One can in principle expect that these answers might be universal in some sense for theories at strong coupling that have a gap in the spectrum of dimensions of operators.

\subsection{Comments on operators generating these metastable states}
\label{sec:operational}

We have translated the increase in energy and angular momentum $E,J$ in the boundary theory due to the bulk excitations on classical stable orbits. However, we have neither specified how to generate these states nor which correlation functions do not thermalize in a typical thermal time in the dual CFT.

\begin{figure}[ht]
\includegraphics[width=7 cm]{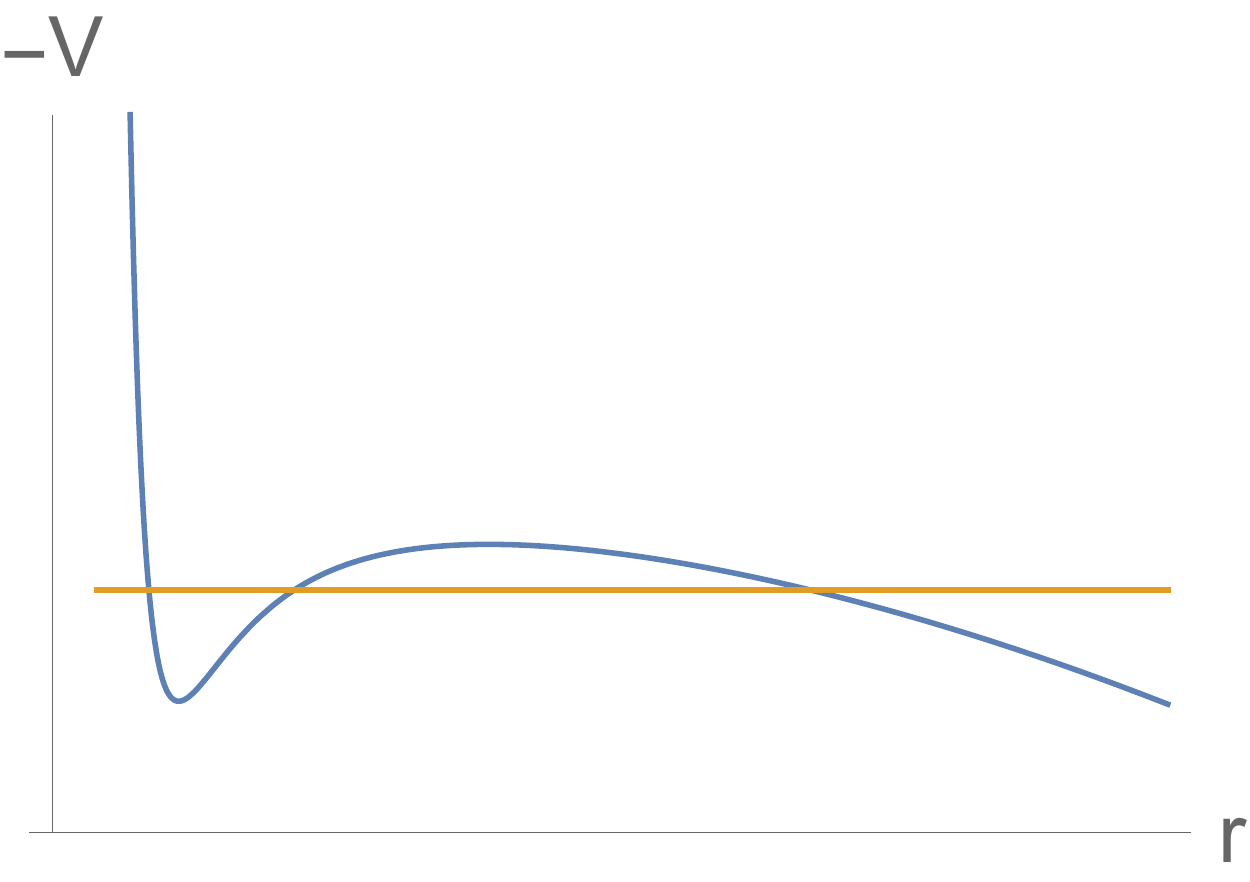} \caption{Effective potential for Euclidean action problem. The horizontal orange line indicates a fixed energy for the motion of the particle.}
\label{fig:potential}
\end{figure}

Two of the authors of this paper recently addressed how to get bulk excitations placing particles in specific global AdS geodesics \cite{Berenstein:2019tcs}. The main idea is that to create particles in the bulk with angular momentum $J$, associated to an operator of dimension $\Delta$, one considers boundary operator insertions of 
\begin{equation}
{\cal O}_{\epsilon, J} \simeq \int d\Omega\, Y_{Jm}(\theta) \exp(-\epsilon H) {\cal O}( \theta, t) \exp(\epsilon H),
\label{eq:boundary-insertion}
\end{equation}
acting on a state representing the background, which in the original paper was the ground state of the conformal field theory. The integrals over the spherical harmonics $Y_{Jm}$ project onto the correct angular momentum state and the presence of $\epsilon H$ regularizes the operator, so that the insertion is normalizable \cite{Berenstein:2014cia}. Because of the Euclidean time evolution on the boundary, the operators in question are not local at an instant of time. The natural way to interpret the preparation of the bulk particle is in terms of tunneling from the boundary. The amplitude to produce the bulk particles can be computed from the Euclidean action of a point particle evaluated on a minimising trajectory connecting the AdS boundary to the turning point (aphelion, $r_\star$) of the classical lorentzian orbit where we want to create the bulk excitation. In this way, the parameter $\epsilon$ becomes the Euclidean time that it takes the euclidean geodesic to connect both points and it can be determined by the properties of the euclidean geodesic solving the matching to the lorentzian orbit. We will be a bit more precise below figure \ref{fig:potential}.

We want to apply this construction to the current discussion, where the state dual to the black hole background is the thermofield double state \cite{Maldacena:2001kr}. The main idea is that this field theory mechanism based on boundary operator insertions allows us to place particles in the classical bulk orbits discussed in the previous section. Moreover, correlation functions of these operators will not only be sensitive to the dynamics of these trajectories, but attuned to them. That is, the physics of these trajectories can be turned to a problem of correlation functions on the boundary. Thus, if we include real time evolution and imaginary time preparation, we at least answer in principle an operational way to access this dynamics. 

To further analyze the interpretation of placing particles of mass $m\simeq \Delta$ in the classical stable geodesics identified in section \ref{sec:bulk}, we need to perform a semiclassical tunneling calculation in  Euclidean AdS. Here, we want to clarify how the result of this euclidean bulk calculation relates to the WKB approximation for the wave function (field) of the particle in the bulk, as discussed in \cite{Festuccia:2008zx}. 

Since our bulk particle has no spin, consider the wave equation for a scalar field $\Phi$ of mass $m$ in the black hole background \eqref{eq:AdsSch}, with angular momentum $J= m \ell$, and energy $E=m e$. Schematically (see appendix \ref{sec:optics} for a more precise derivation), the wave function can be decomposed as
\begin{equation}
  \Phi\simeq f(r, \ell, e) \exp(-i E t ) \exp(i J \phi)\,.
\end{equation}
In a geometric optics approximation, the radial ODE equation for $f(r, \ell, e) \simeq \exp(-m \tilde W(r))$ can be written as
\begin{equation}
  -g^{rr} (\partial_r \tilde W) ^2+ \frac{\ell^2}{r^2} - \frac{e^2}{H(r)}+1=0\,.
\label{eq:constraintWKB}
\end{equation}
The separation of variables in the quantum theory can usually be related to the separation of variables in the Hamilton-Jacobi theory for the classical motion of a particle via the WKB approximation.
In the Hamilton-Jacobi theory, $\tilde W$ is a generating function of a canonical transformation, where the conjugate momentum is $p_r = \partial_r \tilde W$. We can indeed establish this relation by comparing \eqref{eq:constraintWKB} to the constraint equation \eqref{eq:constr}. The match is identical, up to a sign, if we set $g^{rr} p_r = \dot r$, or equivalently, $p_r= g_{rr} \dot r$. This sign corresponds to taking $V(r)\to -V(r), e^2\to -e^2$ and it is standard in the analytic continuation to an Euclidean geometry, as we discuss below and in more technical detail in appendix \ref{sec:euclidean}. 

The euclidean point particle Lagrangian equals
\begin{equation}
L= \frac 12 g_{rr} \dot r^2 + \frac 12 g_{\tau\tau} \dot \tau^2 + \frac 12 g_{\phi\phi} \dot \phi^2,
\end{equation}
where $\tau$ is euclidean time and $g_{\tau\tau}$ is the corresponding euclidean metric component. 
To match the WKB expression for the effective potential, the angular momentum needs to be analytically continued $\ell\to i\ell$, as we analytically continue the Euclidean time. The change of sign of $g_{\tau\tau}$ relative to $g_{tt}$ takes care of the change $e^2\to -e^2$, while the change of sign in the extra term in the potential associated with the constraint equation
arises from requiring that the Hamiltonian constraint has solutions for regular Euclidean geodesics.
All in all, the result changes $e^2-V(r) \to-(e^2 -V(r))$ as is standard in tunneling calculations for a regular Schr\"odinger equation (see appendix \ref{sec:euclidean} for an explicit derivation of the matching between \eqref{eq:constraintWKB} and \eqref{eq:constr}).

If we were to consider a more general generating function
\begin{equation}
W(r,\theta, \tau) =  \tilde W(r) - e \tau + i \ell \phi\,,
\end{equation}
this would correspond to the Hamilton-Jacobi theory generating function $W$ for all the variables, not just the radial direction. This is identical to 
\begin{equation}
W= \int p_r dr -\int e d\tau + \int  \tilde \ell d\phi =  \int p_r \dot r ds-\int e \dot\tau ds + \int \tilde \ell \dot \phi ds= \int  ds 
\label{eq:length}
\end{equation}
where the integral over the worldline  $\int ds$ arises when the constraint equation \eqref{eq:constraintWKB} is applied in the expression. Since $\ell$ is imaginary, the relevant geodesic for the tunneling problem must be complex. In particular, $\phi$ must be imaginary 
so that $W$ remains real. Thus, according to \eqref{eq:length}, evaluation of $W$ computes the length of the geodesic. Hence, $\exp(-2 m W)$ is the probability of tunneling, while $\exp(-m W)$ is the tunneling amplitude itself for the complex trajectory. Notice the mass $m$ of the particle in $AdS$ units plays the role of $1/\hbar$, as discussed when computing the frequency of the radial oscillations for nearly circular stable orbits. By contrast, the tunneling amplitude $\exp(-m\tilde W)$ is the one to be computed when we want to create a particle in one of our classical stable orbits at \emph{fixed} angular momentum. This is because the integration over the $Y_{\ell m}$ in \eqref{eq:boundary-insertion} fixes the angular momentum, on top of the classical energy.  Notice we go from one tunneling expression to the other by a Legendre transform. This modification is also necessary to have a proper variational problem.

Having clarified the relation between the WKB approximation to the bulk wave function and the euclidean action in the inverted potential $-V$, let us have a qualitative discussion on how to implement the ideas from \cite{Berenstein:2019tcs} in the current black hole background situation. The typical inverted potential is shown in figure \ref{fig:potential}, where the yellow line stands for the fixed value of the classical energy.

The prescription in \cite{Berenstein:2019tcs} requires us to evaluate the Euclidean geodesic at fixed $\ell$ (as determined from $Y_{\ell m}$) and between times $\tau= \pm \epsilon$.
The values of $\epsilon$ are obtained by solving for the geodesic once we know the turning point of the trajectory in the inverted potential at fixed energy.
In principle, one then inverts the problem to solve for the value of the energy based on $\epsilon$.
The geodesic will be time symmetric with respect to $\tau=0$, which forces us a to the turning point of the  dynamics in the radial direction at $\tau=0$, namely, to the point where $\dot r=0$.
The geodesic will then need to go to the boundary. The point where the orbit folds back is exactly the aphelion of the regular orbit. Once it is placed at that point, it will oscillate in the radial variable in  real time, starting at that initial condition. 

Given a fixed angular momentum $\ell$, there are three different tunneling problems one may consider depending on the value of the energy $e_\circ$. The first tunneling problem corresponds to the situation where the energy crosses the potential function just once, below the minimum of $-V$. The second corresponds to three crossing points (as the line indicated in the figure). The third corresponds to a single crossing above the maximum of $-V$. 

The prescription in \cite{Berenstein:2019tcs} determines the euclidean time $\epsilon$ as a function of the turning point $r_\star$ once the energy is fixed. To invert this relation gives rise to a multi-valued function $r_\star(\epsilon)$ indicating there is more than one euclidean trajectory solving the problem. However, it is only the trajectory that minimises the euclidean action globally that will dominate the tunneling amplitude.

Remember the maximum of $-V$ corresponds to the circular trajectory in the Lorentzian problem. When the energy is below the minimum of $-V$, the tunneling is to a position of large $r$, i.e. a high energy state in the Lorentzian problem. The corresponding Lorentzian geodesic, starting from $r_\star$ plunges through the horizon and thermalizes. This tunneling problem is very similar to the one discussed in global $AdS$ \cite{Berenstein:2019tcs}. Hence, it should dominate at small $\epsilon$. For the second type of energy, the tunneling is to a stable orbit with some radial motion, starting from the outermost $r_\star$. 

As the energy approaches the local maximum of $-V$, the proper Euclidean time diverges since the trajectory spends a very long time near the top of the potential. This means that exactly for the circular orbit we have that $\epsilon(e_\circ, \ell)\to \infty$. Beyond this energy, the time coordinate $\epsilon$ needs to come back down again and be finite. There would also be a discontinuous jump in $r_\star$ to a small value. These geodesics would have the same $\epsilon$ as other geodesics that have high energy. That is, if we invert the problem to find $e(\epsilon,\ell)$, the energy $e$ is multivalued.
Basically, for each such value of $\epsilon$ there is more than one Euclidean geodesic
that can contribute. The one that dominates is the one that ends at the largest value of $r$, just like in vacuum AdS, because it has the shortest length. This should be the one with high energy, which is most similar to vacuum AdS.
Effectively, our prescription with $\epsilon$ will not let us probe the small $r$ region with plunging geodesics that start closer to the origin than the circular orbit: they correspond to a subdominant saddle in the path integral.
Many of these will have $e<\ell$. This is very similar to results in \cite{Leichenauer:2013kaa}, which argue that 
certain smearing functions can not be constructed for situations where the energy is small relative to the angular momentum  $e\ll \ell$. 

In our setup, the full Euclidean calculation should not use just the naive analytic continuation of the time variable, but should  also include the temperature of the black hole, with a ``cigar-like" geometry. The time variable should be periodic.
That is, we usually need to impose regularity at the horizon. The periodicity  of the circle at infinity is the temperature of the black hole. This would force an upper bound of $\epsilon<\beta/4$, because we need to identify $\tau\to \tau +n \beta$. The physics of this is as follows: if we have too long of a wait in Euclidean time, we will be sensitive to thermal fluctuations and this will take the particle away from the trajectory we desire. These thermal fluctuations are calculated from the geodesics with the different values of $   \tau\to \tau +n \beta $. The bound guarantees the geodesic we focus on is the dominant one. 
One should be careful of this analysis in the case of microstate geometries, rather than in the canonical ensemble.
There the periodicity in Euclidean time is not guaranteed and this can make its presence known in Euclidean correlators \cite{Chen:2018qzm}.

To summarize, there can be many tunneling Euclidean geodesics subtended from fixed points in the boundary. 
Only one such geodesic will be dominant and these dominant geodesics  miss regions of the interior where in principle  excitations can be placed in the bulk.

For completeness, we can also consider the analysis of the oscillating region within a WKB approximation. These will give the real part of the exact quantized energies of the particles in the stable orbits. The orbit in the radial direction would be periodic in time, and one can semiclassically quantize 
the energy by requiring that
\begin{equation}
\oint p_r dr = \left(n+\frac 12\right) 2\pi \hbar 
\label{eq:WKBcond}
\end{equation} 
where the contour integral covers the physical region twice. Here one can take $\hbar= 1/m$, or take $\hbar =1$, but use a $p_r$ that includes the factor of $m$ derived from the correct normalization of the action. This is the type of analysis that was performed in \cite{Festuccia:2008zx}, see also the earlier work \cite{Grain:2006dg}. 
For small oscillations near the circular orbit, this is already implicit in our analysis \eqref{eq:desc}. That analysis quantized a harmonic oscillator with the frequency of oscillations of the radial variable near the circular orbit. A more general analysis would compute the contour integrals and impose the condition \eqref{eq:WKBcond} to evaluate the energies of the orbits.

\section{Outlook}
\label{sec:outlook}

In this paper we took some steps to understand the dual interpretation to 
the physics of stable geodesics in black hole backgrounds in AdS/CFT.  
We argued their existence indicates that there is some physics that does not thermalize on a thermal time scale and we interpreted the bulk excitations as metastable states in the dual field theory. We studied the size, angular velocity, energy and angular momentum of these trajectories. Using the standard AdS/CFT dictionary, the last two provide the quantum numbers of these states. The least energetic marginally stable trajectory, the ISCO, provides a sharp way to identify the window of quantum numbers where this type of dynamics exists.

We were careful to compute the spectrum of these excitations near the circular stable geodesics, including the radial excitation quantum numbers. The results indicate a binding energy for both the radial fluctuations and for the circular trajectories, which usually vanishes at very large angular momentum. Such a small correction is not compatible with a free field theory analysis, where the spectrum of excitations has an integer spacing between them. We do not yet understand under what conditions these results are universal. However, using holographic considerations would suggest the dual field theory to have a gap in the spectrum and to be sparse so that the gravity dual might be a good approximation to the physics \cite{Heemskerk:2009pn,Hartman:2014oaa}.

In the large temperature limit, the CFT predictions on a sphere or on the plane should be equivalent, up to finite volume effects. Since the energy of the bulk excitations scales differently with the temperature than linearly, it requires a second scale in the problem. We argued this scale is the curvature of the boundary field theory since these orbits do not exist when the boundary is flat, like in BTZ black holes, nor in flat AdS black holes.

The positive curvature of the sphere in the boundary is suggestive. Light rays emanating from a point
focus and might reconstitute objects when they do. Since we don't have a CFT calculation to perform, we can guess that this focusing is an important ingredient in the corresponding calculation. Such an idea suggests that for negatively curved boundaries the corresponding black holes do not have such stable trajectories, as the null geodesics defocus. 
 These also have an important role to play in the understanding of entanglement
 entropy \cite{Casini:2011kv}. This analysis is beyond the scope of the present paper.

At least in principle, our analysis provides a probe/test of which strongly coupled theories are holographic: they need to have these long lived excitations. Because these only occur when the boundary has spatial curvature, we need a 2d material sample in the shape of a sphere to use as a probe.
The quantum numbers of these excitations depend on an anomalous scaling with the temperature and occur at non-zero angular momentum. Alternatively, we can fix the temperature and change the size of the sample: their energies would depend 
non-trivially on the radius of the sphere. 
Another probe that depends on the geodesic structure is the light ring (Einstein ring). This has been argued 
to be probed by CFT correlators that see gravitational lensing in \cite{Hashimoto:2018okj}. In our results, the light ring radius also has an anomalous scaling at high temperatures that is different than the horizon radius.

Our work focused on spherically symmetric AdS black holes. It would obviously be interesting to extend these calculations to other black holes in AdS, like Reissner-Nordstrom or Kerr, whose structure will necessarily be richer.

We also discussed a CFT mechanism to prepare the bulk particles on these trajectories. This involves tunneling from the boundary using euclidean geodesics. In general, these tunneling amplitudes are hard to compute for these black holes, as they require solving complicated integrals. It would be interesting to find cases where some of these can be performed analytically. Since these tunneling geodesics to the boundary have infinite action, to understand their physics better one needs to find a regularization scheme that can be used to compare directly with the physics of empty AdS. Having explicit analytical examples would provide a guide to do so.  We also analyzed in more detail the relation between this tunneling computation and the WKB approximation for the radial part of the scalar wave equations for massive particles in AdS. 

Another important observation we have is that there can be more than one tunneling geodesic related to fixed end points and angular momentum on the boundary. This is related to the Euclidean time being periodic in the Euclidean black hole background. The bulk geodesics attached to these points can have different winding numbers.  
These make it hard to prepare states in certain 
regions of the black hole geometry with our prescription: near the horizon at large angular momentum, as they correspond to a subdominant contribution in the tunneling trajectory. One can speculate that these obstructions are related to the absence of smearing functions for modes very close to the black hole with large angular momentum \cite{Leichenauer:2013kaa}, which seem somewhat similar in character.

\acknowledgements

D. B. would like to thank D. Harlow, J. Maldacena, J. Santos for discussions. 
The work of D.B. is supported in part by the Department of Energy under grant {DE-SC} 0011702. 


\appendix

\section{Geometric optics approximation}
\label{sec:optics}

On general grounds, one expects our geodesic analysis in section \ref{sec:bulk} to capture the geometric optics limit of a massive scalar field propagating in \eqref{eq:AdsSch}. We make this connection explicit below, explaining why the temperature scaling in \eqref{eq:scaling} was observed in \cite{Festuccia:2008zx} when studying the temperature dependence on the quasi-normal modes propagating in \eqref{eq:AdsSch}. This discussion will also allow us to connect with our euclidean tunneling observations in the main text.

Consider a massive bulk scalar field propagating in \eqref{eq:AdsSch} with action
\begin{equation}
    S = -\frac{1}{2} \int d^{d+1}x\,\sqrt{-g}\left((\nabla\Phi)^2 + m^2\Phi^2\right)
\end{equation}
Decomposing the scalar wave function as $\Phi = e^{-i\omega t}\,Y_{lm}(\vec{\theta})\,r^{-(d-1)/2}\,\psi(r)$, its Klein-Gordon equation of motion reduces to the radial equation \cite{Festuccia:2008zx}
\begin{equation}
    \left(\partial_z^2 + V_\Phi - \omega^2\right)\psi = 0\,, \quad\quad V_\Phi = H(r)\left(\frac{(2l+d-2)^2+1}{4r^2} + \nu^2-\frac{1}{4} + 2M \frac{(d-1)^2}{4r^d}\right)\,.
\label{eq:KG-wave}
\end{equation}
The parameter $\nu^2 = (mL)^2 + d^2/4$, i.e. the conformal dimension of the boundary operator $\mathcal{O}$ is $\Delta = \nu + d/2$, and the radial coordinate $z$ is defined through
\begin{equation}
 \frac{dz}{dr} = -\frac{1}{H(r)}\,.
\end{equation}

Since the conserved charges \eqref{eq:conserved} were per unit mass, the relation between these and the Fourier modes of the scalar field must be 
\begin{equation}
  \omega = mL\,e \quad \quad l = mL\,\ell
\end{equation}
The geometric optics limit requires to work with a large mass, so that $\nu \approx mL$. Notice the last term in the potential $V_\Phi$ is negligible both for large orbits at finite $M$ and in the large $M$ limit keeping $M/r^{d-2}$ fixed. Both situations involve large angular momentum, so that we can approximate the scalar potential by
\begin{equation}
    V_\Phi \approx \nu^2\,H(r)\left(1+\frac{\ell^2}{r^2}\right)=\nu^2\,V_\ell(r)\,.
\end{equation}
Looking for a radial wave function of the form $\psi(r) = e^{\nu S}$ and only keeping the dominant contribution to the wave equation \eqref{eq:KG-wave}, i.e. keeping terms not suppressed by negative powers of $\nu$, the wave equation reduces to
\begin{equation}
    -(\partial_z S)^2 + V_\ell = e^2\,.
\end{equation}
This equation reproduces the geodesic constraint \eqref{eq:constr} if $S = i\tilde{W}$ and we use the Hamilton-Jacobi equation $p_r=\partial_r\tilde{W}$. Indeed,
\begin{equation}
   -(\partial_z S)^2=-(H(r))^2(\partial_r S)^2 =H^2\,(\partial_r \tilde{W})^2= p_r^2\,H^2 = H^2\,(g_{rr}\dot{r})^2 = \dot{r}^2
\end{equation}
We learn the radial wave function of the scalar field equals $\psi(r) = e^{i\nu\tilde{W}}$ in the geometric optics limit, where $\tilde{W}$ stands for the lorentzian particle action in the Hamilton-Jacobi formulation, as claimed in the main text.

\subsection{Euclidean continuation}
\label{sec:euclidean}

The discussion above related the radial wave equation of the massive scalar field with the geodesic constraint equation \eqref{eq:constr}. Here, we discuss the euclidean continuation of the lorentzian geodesic action in order to compute the amplitude for tunneling of a bulk particle as the interpretation of the boundary operator insertion \eqref{eq:boundary-insertion}.

To avoid any confusions regarding our starting action \eqref{eq:geodesic-action} and our conventions in the main text, remember the standard lorentzian relativistic point particle lagrangia
\begin{equation}
    L_\mt{lor} = -m\sqrt{-\dot{x}^2}
\end{equation}
is classically equivalent to
\begin{equation}
  L_\mt{lor} = m\left(\frac{\dot{x}^2}{2em} - \frac{em}{2}\right) = m\hat{L}_\mt{lor} + \text{constant}
\end{equation}
Fixing the gauge $em=1$ gives rise to our starting lorentzian action \eqref{eq:geodesic-action} together with the constraint \eqref{eq:constraint}. This last step, where we dropped the constant piece, justifies why the conserved charges \eqref{eq:conserved} were defined per unit of mass. 

One possible euclidean continuation of the Lorentzian lagrangian
\begin{equation}
    \hat{L}_\mt{lor} = \frac{1}{2}\left(g_{tt}\dot{t}^2 + g_{rr}\dot{r}^2 + g_{\phi\phi}\dot{\phi}^2\right)
\end{equation}
is given by $t\to i\tau$ and $s\to -is$, where $s$ is proper time. The change in the measure $ds$ together with an overall sign change in the lagrangian density allows to transform the amplitude in the lorentzian path integral  $e^{iS_\mt{lor}}$ into $e^{-mS_\mt{euc}}$ with corresponding euclidean lagrangian given by
\begin{equation}
    L_\mt{euc} = \frac{1}{2}\left(-g_{tt}\dot{\tau}^2 +g_{rr}\dot{r}^2 + g_{\phi\phi}\dot{\phi}^2\right) \equiv \frac{1}{2}\left(g_{\tau\tau}\dot{\tau}^2 + g_{rr}\dot{r}^2 + g_{\phi\phi}\dot{\phi}^2\right)
\end{equation}
where $g_{\tau\tau}=-g_{tt}$ stands for the standard euclidean time component of the metric and $L_\mt{euc}$ is based on the euclidean metric, as it should\footnote{More explicitly, what has happened is that $L_\mt{lor}\to -L_\mt{euc}$ and this sign is absorbed in the amplitude $e^{-mS_\mt{euc}}$ together with the change in the measure induced by $ds\to -ids$.}.

Let us check what happens to the constraint equation \eqref{eq:constraint} under this euclidean continuation
\begin{equation}
    -1 = \frac{e^2}{g_{tt}} - g_{rr}\dot{r}^2 - g_{\phi\phi}\dot{\phi}_\mt{e}^2
\end{equation}
Since both $t$ and $s$ are Wick rotated, $e$ does not change, but the kinetic terms in the spatial directions flip sign. Writing this in terms of the euclidean metric components, this is equivalent to
\begin{equation}
  \frac{e^2}{g_{\tau\tau}} = -g_{rr}\dot{r}^2 + 1 - g_{\phi\phi}\dot{\phi}_\mt{e}^2  
\end{equation}
Finally, $\dot{\phi}=-i\dot{\phi}_\mt{e}$. This requires the Wick rotation of the angular momentum $\ell \to i\ell$, allowing us to write the euclidean geodesic constraint as
\begin{equation}
  \frac{e^2}{g_{\tau\tau}} = -g_{rr}\dot{r}^2 + 1 + \frac{\ell^2}{r^2}
\end{equation}
This is the same constraint equation as in \eqref{eq:constraint} but with a negative sign in the radial kinetic term. This is the right equation to perform the tunneling calculation in the point particle (WKB) approximation to the bulk scalar field amplitude. We can indeed get to this equation by $\psi=e^{-\nu \tilde{W}}$, as used in the main text.

\end{document}